\documentclass[10pt,conference]{IEEEtran}
\renewcommand{\baselinestretch}{1.}
\usepackage{subcaption}
\usepackage{ifpdf}
\ifCLASSINFOpdf
\usepackage{graphicx}
\usepackage{grffile}
\usepackage{epstopdf}
\usepackage{grffile}
\usepackage[export]{adjustbox}
\else
\fi
\usepackage[cmex10]{amsmath}
\usepackage{booktabs}
\usepackage{array}
\usepackage{amsfonts}
\usepackage{lipsum}
\usepackage{multicol}
\usepackage{mathtools}
\DeclarePairedDelimiter\ceil{\lceil}{\rceil}
\DeclarePairedDelimiter\floor{\lfloor}{\rfloor}
\usepackage[numbers,sort&compress]{natbib}
\usepackage{bm}
\makeatletter

\newcommand{\Rmnum}[1]{\expandafter\@slowromancap\romannumeral #1@}
\makeatother

\begin{document}
\title{Thinned Coprime Arrays for DOA Estimation}
\author{\IEEEauthorblockN{Ahsan Raza\IEEEauthorrefmark{1}, Wei Liu\IEEEauthorrefmark{1} and Qing Shen\IEEEauthorrefmark{2}}\\
\IEEEauthorblockA{\IEEEauthorrefmark{1}Communications Research Group}
{Department of Electronic and Electrical Engineering, University of Sheffield}\\
{Sheffield, S1 3JD, U.K.}\\
\IEEEauthorblockA{\IEEEauthorrefmark{2}School of Information and Electronics, Beijing Institute of Technology}
{Beijing, 100081, China}\\

  }

\maketitle


\begin{abstract}
Sparse arrays can generate a larger aperture than traditional uniform linear arrays (ULA) and offer enhanced degrees-of-freedom (DOFs) which can be exploited in both beamforming and direction-of-arrival (DOA) estimation. One class of sparse arrays is the coprime array, composed of two uniform linear subarrays which yield an effective difference co-array with higher number of DOFs. In this work, we present a new coprime array structure termed thinned coprime array (TCA), which exploits the redundancy in the structure of the existing coprime array and achieves the same virtual aperture and DOFs as the conventional coprime array with much fewer number of sensors.  An analysis of the DOFs provided by the new structure in comparison with other sparse arrays is provided and simulation results for DOA estimation using the compressive sensing based method are provided.
\end{abstract}


\IEEEpeerreviewmaketitle


\section{INTRODUCTION}


Sparse arrays can detect more sources than the number of sensors due to increased number of degrees of freedom (DOFs) available from their difference co-array model~\cite{pillai89,vantrees02a}. These DOFs represent the different lags at which the autocorrelation can be computed from the data. Different types of sparse arrays have been proposed recently. Minimum redundancy array (MRA) is a sparse array that maximizes the number of consecutive lags in the difference co-array for a fixed number of sensors \cite{moffet68a}. Significant contribution to MRA for a large number of antennas was presented in \cite{ishiguro80}. Another sparse array termed as minimum hole array (MHA) minimizes the number of holes in the difference co-array \cite{golomb77}. However, MRA and MHA do not have closed-form expressions for the array geometry and the sensor positions are normally extracted from tabulated entries~\cite{moffet68a}.

Nested arrays are sparse arrays composed of two uniform linear subarrays where one subarray is denser with unit inter-element spacing than the other one~\cite{pal10}. It has the ability to resolve $O$($N^{2}$) sources with $N$ sensors. In comparison to MRA and MHA, nested array is simple to construct and exact expressions are available for sensor locations and computing DOFs for a given number of sensors. Two-dimensional extensions of nested arrays were also provided in \cite{PalV12a, PalV12b}. Nested arrays possess hole-free co-arrays which gives them an edge in their DOA estimation performance, but due to a densely packed subarray, they are prone to the effect of mutual coupling~\cite{gupta83}. Coprime arrays are sparse arrays composed of two uniform linear subarrays where one subarray has $M$ sensors with $Nd$ inter-element spacing, while the other subarray has $N$ sensors with $Md$ inter-element spacing where $M$ and $N$ are coprime integers and $d$ is the unit spacing set to be $\frac{\lambda}{2}$ with $\lambda$ corresponding to the wavelength of the impinging signal. This structure is referred as the prototype coprime array with $M+N-$1 sensors \cite{vaidyanathan11a} and provides 2($M$+$N$)-1 consecutive lags. A modification to this coprime array structure was proposed in \cite{pal11a} by increasing the number of elements in one subarray from $M$ sensors to 2$M$ sensors. This structure of $2M+N-1$ sensors termed as conventional coprime array resulted in a significant increase in consecutive lags by providing $2MN+2M-1$ consecutive lags which can be exploited using subspace based DOA estimation methods such as MUSIC~\cite{schmidt86a,pal11a,liu15remarks}.

Two generalized coprime array configurations were recently proposed in \cite{qin15a}, where the first type was based on compressing the inter-element spacing of the $N$-element subarray by factors of $M$, resulting in a coprime array with compressed inter-element spacing (CACIS). The minimum inter-element spacing in CACIS remains unit spacing with considerable overlapping between self lags and cross lags. To counter this, a second type of array was proposed by introducing displacement between the two subarrays, resulting in an array with a larger minimum inter-element spacing, larger aperture and higher number of unique lags. This array was termed as coprime array with displaced subarrays (CADiS). It is shown that the CADiS structure yields the highest number of unique lags which can all be exploited using compressive sensing (CS) based DOA estimation methods.

In this paper, we propose a thinned coprime array (TCA), a new structure resulting from exploiting the redundancy in the difference co-array model of the conventional coprime array. As proved later in the paper, the lag contribution from some of the sensors in the 2$M$-element subarray of the conventional coprime array is generated by the rest of the sensors in the array and these sensors can therefore be removed without affecting the properties of the parent array. The proposed TCA holds the same number of consecutive lags, unique lags and aperture as the conventional coprime array with  $\ceil*{\frac{M}{2}}$ fewer sensors. In comparison to other sparse arrays such as the nested array, for a fixed number of sensors, the thinned coprime array achieves unique lags more than the hole-free structure of the nested array, contains significant number of consecutive lags, produces a much larger aperture and provides a much sparser array structure than the nested array.

This paper is organized as follows. The conventional coprime array model is reviewed in Section \ref{sec:spmodel}. The redundancy in the conventional coprime array is analyzed with a detailed proof and the new TCA structure is proposed in Section \ref{sec:tcsa}.  A comparison in terms of DOFs between the TCA and other sparse arrays is provided in Section \ref{sec:DOF}. Simulations results using compressive sensing (CS) based DOA estimation method is provided in Section \ref{sec:doa}, followed by conclusions drawn in Section \ref{sec:conclusion}.


\section{Conventional coprime array}
\label{sec:spmodel}
For the conventional coprime array with $2M+N-1$ sensors, where $M$ and $N$ are coprime integers, the array sensors are positioned at
\begin{equation}
\mathbb{P}=\{Mnd\mid\text{$0 \leq n \leq N-1$}\} \cup \{Nmd\mid\text{$0 \leq m \leq 2M-1$}\}
\end{equation}
\begin{figure}
\centering
\includegraphics[width=.5\textwidth]{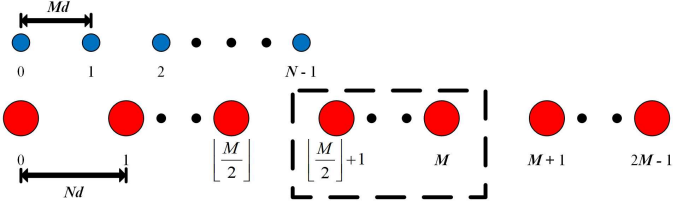}
\caption{Conventional coprime array}
\label{fig:csa}
\end{figure}
The positions of the sensors are given by the set $\textbf{p}=[p_1,...p_{2M+N-1}]^T$ where $p_i \in \mathbb{P}, i=1,...2M+N-1$. The first sensor in both subarrays is co-located at the  zeroth position with $p_1=0$. 

Consider the scenario where $Q$ uncorrelated signals are impinging on the array from angles $\Theta=[\theta_1, \theta_2,...\theta_Q]$ and their sampled baseband waveforms are expressed as $s_q(t), t=1, ..., T$, for $q=1, ..., Q$. Then, the data vector received by the coprime array is given by
\begin{equation}
\textbf{x}(t)=\sum_{q=1}^{Q} \textbf{a}(\theta_q) s_q(t)+\textbf{n}(t) = \textbf{A}\textbf{s}(t)+ \textbf{n}(t)
\end{equation}
where
\begin{equation}
\textbf{a}(\theta_q)=[1,e^{-j \frac{2\pi p_2}{\lambda} sin(\theta_q)},....,e^{-j \frac{2\pi p_{2M+N-1}}{\lambda} sin(\theta_q)}]^T
\end{equation}
is the steering vector of the array corresponding to $\theta_q$, $\textbf{A}=[\textbf{a}(\theta_1),...,\textbf{a}(\theta_Q)]$ and $\textbf{s}(t)=[s_1(t),...s_Q(t)]^T$. The entries of the noise vector $\textbf{n}(t)$ are assumed as independent and identically distributed (i.i.d) random variables following a complex Gaussian distribution $CN(0,\sigma_n^2\textbf{I}_{2M+N-1})$. The covariance matrix of data vector $\textbf{x}(t)$ is given by
\begin{equation}
\textbf{R}_{xx}=E[\textbf{x}(t) \textbf{x}^H(t)]=\textbf{A}\textbf{R}_{ss}\textbf{A}^H+\sigma_n^2\textbf{I}_{2M+N-1}
\end{equation}
\begin{equation}
\textbf{R}_{xx}=\sum_{q=1}^{Q}\sigma_q^2\textbf{a}(\theta_q)\textbf{a}^H(\theta_q)+\sigma_n^2\textbf{I}_{2M+N-1}
\end{equation}
where $\textbf{R}_{ss}=E[\textbf{s}(t) \textbf{s}^H(t)]=$ diag$([\sigma^2_1,...,\sigma^2_Q])$ is the source covariance matrix, with $\sigma^2_Q$ denoting the signal power of the $q$th source. In practice, the covariance matrix is estimated from the $T$ available samples.
\begin{equation}
\textbf{R}_{xx}=\frac{1}{T}\sum_{t=1}^{T}[\textbf{x}(t) \textbf{x}^H(t)]
\label{eq:Covar}
\end{equation}
From the antennas located at the $m$th and $n$th positions in $\textbf{p}$, the correlation $E[\textbf{x}_m(t)\textbf{x}^*_n(t)]$ results in the $(m,n)$th entry in $\textbf{R}_{xx}$ with lag $p_m-p_n$. All the values of $m$ and $n$, where $0 \leq m,n \leq 2M+N-1$, yield the lags or virtual sensors of the following difference co-array:
\begin{equation}
\mathbb{C_P}=\{\textbf{z}\mid\text{$\textbf{z} = \textbf{u}-\textbf{v}, \textbf{u} \in \mathbb{P}, \textbf{v} \in \mathbb{P}$}\}\;.
\end{equation}


\section {THINNED COPRIME SENSOR ARRAY}
\label{sec:tcsa}
\label{sec:tcsa_csa}
Conventional coprime arrays yield consecutive lags from $-MN$ to $MN$ for a given $M$ and $N$. In this section we will show that some of the sensors in the $2M$-element subarray as depicted by dashed rectangles in Fig. \ref{fig:csa} are redundant as their contribution of lags is generated by the rest of the sensors in the array and they can be removed to yield the proposed thinned coprime array.

\textbf{Theorem}\textbf{.} \textit{The number of redundant sensors in a conventional coprime array with $M \geq $ 4 and $N \geq $ 5 are given by}
\begin{equation}
S_{red} = \ceil*{\frac{M}{2}}
\label{eq:csa_red}
\end{equation}
\textit{where the starting index of these $S_{red}$ contiguous redundant sensors in the }(2$M-$1)-\textit{element subarray is given by} $\floor*{\frac{M}{2}} + $1.
\begin{IEEEproof} The structure of the difference co-array can be divided into \textit{self difference} i.e. diff($\mathbb{A}$, $\mathbb{A}$) and diff($\mathbb{B}$, $\mathbb{B}$) and \textit{cross difference} i.e.  diff($\mathbb{A}$, $\mathbb{B}$) and  diff($\mathbb{B}$, $\mathbb{A}$) where $\mathbb{A}$ and $\mathbb{B}$ contain the sensor positions $M n$ and $N m$ respectively for the two subarrays with $0 \leq n \leq N - $ 1 and $0 \leq m \leq$ 2$M - $ 1. The self difference sets diff($\mathbb{A}$, $\mathbb{A}$) and diff($\mathbb{B}$, $\mathbb{B}$) are given by
\begin{align*}
 & \{Mn_1 - Mn_2 \mid\text{$0 \leq n_1, n_2 \leq N-1$}\},\\
 & \{Nm_1 - Nm_2 \mid\text{$0 \leq m_1, m_2 \leq 2 M - 1$}\}.
\end{align*}
 while the cross difference sets diff($\mathbb{A}$, $\mathbb{B}$) and diff($\mathbb{B}$, $\mathbb{A}$) are given by
 \begin{align*}
  \{\pm (Mn - Nm) \mid\text{$0 \leq n \leq N - 1, 0 \leq m \leq 2M-1$}\}.
\end{align*}

We start with the scenario where $M$ is even and $N$ is odd for the structure in Fig. \ref{fig:csa} and generate the cross difference diff($\mathbb{A}$, $\mathbb{B}$) matrix with respective index ($n$, $m$) corresponding to the lag entry $Mn - Nm$. It was shown in \cite{shen15} that the entries of cross correlation matrix with indices ($n_1$, $m_1$) and ($n_2$, $m_2$) were found to be complex conjugate of each other when the indices satisfied the following relationship
\begin{equation}
( n_1 + n_2 ) M = (m_1 + m_2) N
\label{eq:comp_conjugate}
\end{equation}
with the sufficient condition for \eqref{eq:comp_conjugate} given by
\begin{equation}
(n_1 + n_2  = N) \cap (m_1 + m_2 = M)
\label{eq:condition_comp}
\end{equation}

For cross difference matrix diff($\mathbb{A}$, $\mathbb{B}$), this condition dictates that if we consider an index ($n_1$, $m_1$) with $m_1$ in the range 0 $\leq m_1 \leq \floor*{\frac{M}{2}} - $1 (for even $M$, $\floor*{\frac{M}{2}} - $1 changes to ${\frac{M}{2}} - $1) and $n_1$ from 1 $\leq n_1 \leq N - $1, then it will have a corresponding index ($n_2$, $m_2$) i.e. ($N - n_1$, $M - m_1$) with $m_2$ in the range $\frac{M}{2}+ 1 \leq m_2 \leq M$ (for even $M$, ${\frac{M}{2}} +$1 is the same as $\floor*{\frac{M}{2}}+ $1) and $n_2$ from 1 $\leq  n_2 \leq N - $1 with both indices satisfying \eqref{eq:comp_conjugate}. The corresponding entries of cross difference matrix with indices ($n_1$, $m_1$) and ($n_2$, $m_2$) satisfy the following relationship.
\begin{eqnarray}
\text{diff}(\mathbb{A}, \mathbb{B})^{n_1, m_1} &=& - \text{diff}(\mathbb{A}, \mathbb{B})^{n_2, m_2} \nonumber \\
&=& - \text{diff}(\mathbb{A}, \mathbb{B})^{N - n_1, M - m_1}
\label{eq:rel_comp}
\end{eqnarray}

It thus follows that the lag entries corresponding to index range ($n_2$, $m_2$) of diff($\mathbb{A}$, $\mathbb{B}$) will all be found in lag entries corresponding to index range ($n_1$, $m_1$) of diff($\mathbb{B}$, $\mathbb{A}$) making the contribution of these lags from  index ($n_2$, $m_2$) redundant.

For index ($n_1$, $m_1$) with $m_1 = \floor*{\frac{M}{2}} = {\frac{M}{2}}$, the corresponding index ($n_2$, $m_2$) where 1 $\leq n_1,n_2 \leq N - $1, will also have $m_2 = \frac{M}{2}$  with indices and their entries satisfying \eqref{eq:comp_conjugate} and \eqref{eq:rel_comp} respectively in the same column. 

As for the lag entries $-Nm$ with index range (0, $m$) where $n = $ 0 and $\frac{M}{2}+ 1 \leq m \leq M$, we consider index $m'$  where (0 $\leq m' \leq \frac{M}{2} $) $ \cup $ ($M+ $ 1 $\leq m' \leq $ 2$M - $ 1). Then, by taking self difference diff($\mathbb{B'}$, $\mathbb{B'}$)  where $\mathbb{B'}$ contains entries $Nm'$, the lag entries for index range (0, $m$) can all be generated. As all the lags of sensors in the (2$M - $1)-element subarray positioned at ($\frac{M}{2}+ $ 1) $N \leq mN \leq MN$ have been generated by the remaining sensors in the array, it proves the existence of $\ceil*{\frac{M}{2}}$ redundant sensors shown by dashed rectangle in Fig. \ref{fig:csa}.

For the scenario where $M$ is odd and $N$ is even, we again generate the cross difference matrix diff($\mathbb{A}$, $\mathbb{B}$). Then by considering an index ($n_1$, $m_1$) with $m_1$ in the range 0 $\leq m_1 \leq \floor*{\frac{M}{2}}$ (for odd $M$, $\floor*{\frac{M}{2}}$ changes to ${\frac{M - 1}{2}}$) and $n_1$ from 1 $\leq n_1 \leq N - $1, it will have a corresponding index ($n_2$, $m_2$) i.e. ($N - n_1$, $M - m_1$) with $m_2$ in the range $\frac{M+1}{2} \leq m_2 \leq M$ (for odd $M$, ${\frac{M+1}{2}}$ is the same as $\floor*{\frac{M}{2}}+ $1)  and $n_2$ from 1 $\leq n_2 \leq N - $1 with both indices satisifying \eqref{eq:comp_conjugate}. The corresponding entries of index ($n_1$, $m_1$) and ($n_2$, $m_2$) respectively satisfy \eqref{eq:rel_comp}. 

It follows that the lag entries corresponding to index range ($n_2$, $m_2$) of diff($\mathbb{A}$, $\mathbb{B}$) will all be found in lag entries corresponding to index range ($n_1$, $m_1$) of diff($\mathbb{B}$, $\mathbb{A}$). As for the lag entries $-Nm$ with index range (0, $m$) where $n = $ 0 and $\frac{M+1}{2} \leq m \leq M$, we consider index $m'$  where (0 $\leq m' \leq \frac{M - 1}{2} $) $ \cup $ ($M+ $ 1 $\leq m' \leq $ 2$M - $1), then by taking self difference diff($\mathbb{B'}$, $\mathbb{B'}$)  where $\mathbb{B'}$ contains entries $Nm'$, the lag entries for index range (0, $m$) can all be generated.  As all the lags of sensors in (2$M-$1)-element subarray positioned at ($\frac{M+1}{2}N$) $\leq mN \leq MN$ have been generated by rest of the sensors in the array, it again proves the existence of $\ceil*{\frac{M}{2}}$ or ${\frac{M + 1}{2}}$ redundant sensors shown by dashed rectangle in Fig. \ref{fig:csa}. The proof is equally applicable for the case of both odd valued $M$ and $N$.
\end{IEEEproof}


\section{DOF COMPARISON OF SPARSE ARRAYS}
\label{sec:DOF}

In this section we compare the number of DOFs provided by the proposed thinned coprime array to nested arrays, CADiS and its special cases for a fixed number of total sensors in the array.

Nested arrays for a given $N_1$ and $N_2$, where $N_1$ and $N_2$ represent the number of sensors in the two constituent subarrays, provide a hole free coarray of 2$N_2$($N_1 +$ 1)$- $1 lags for a total of $N_1 + N_2$ sensors in the array. The CADiS structure in \cite{qin15a} brings two changes to the existing prototype coprime array. In the first change, the first subarray of $N$ sensors which originally has an interelement spacing of $Md$ is compressed by a factor $p$ where we assume $M = pM'$ for some $p$ that takes value in the range 2 $\leq p \leq  M$ with $1 \leq M' < M$ ($M' = $1 is a special case for nested CADiS which will be discussed later). The resulting factor $M'$ and $N$ are still coprime. The elements of the first subarray then possess an interelement spacing of $M'd$ while the second subarray of $M$ sensors retains the original interelement spacing of $Nd$.

For the second change, it displaces the two subarrays by a factor $Ld$ which ensures a larger minimum interelement spacing and increased number of unique lags. It was shown in \cite{qin15a} that the CADiS configuration for $M' > $ 1 achieves a maximum number of unique lags equal to 2$MN + $2$M - $5 when $L > N$($M - $2), while the maximum number of consecutive lags are achieved when $L = M' + N$ with $MN - $($M' - $1)($N - $2)$ + $1 consecutive lags and 2$MN + $2$M' - $1 unique lags. The number of unique lags increase with increasing $M'$ while the consecutive lags decrease. Nested CADiS with $M' = $1 provides a hole-free co-array of 2$MN + $1 lags.
\begin{figure}[htbp]
\begin{subfigure}[b]{0.245\textwidth}  \includegraphics[width=\linewidth]{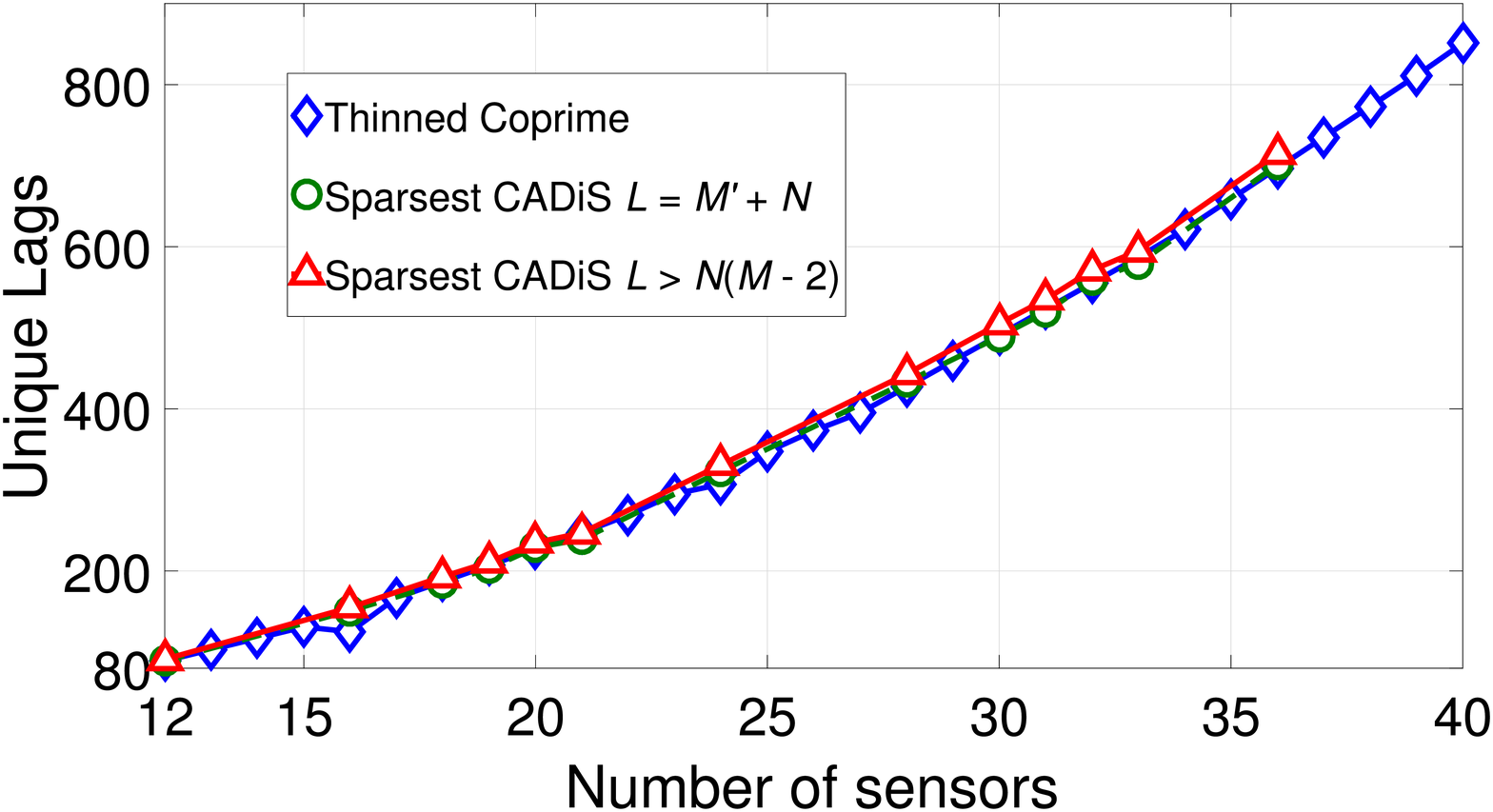}    \caption{Unique lags}    \label{fig:f1}  \end{subfigure}%
\begin{subfigure}[b]{0.245\textwidth}  \includegraphics[width=\linewidth]{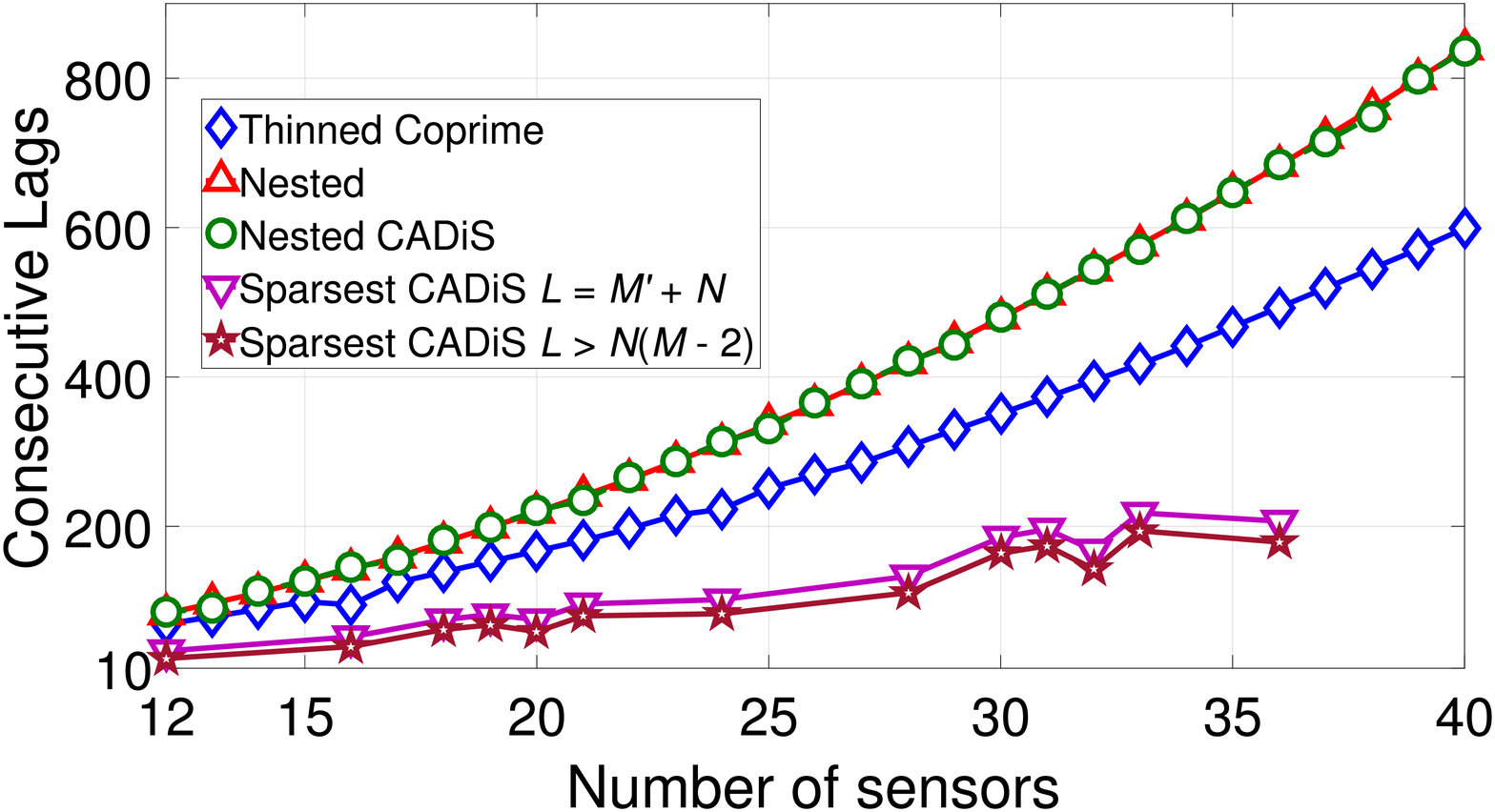}   \caption{Consecutive lags}      \label{fig:f2}   \end{subfigure}
 \caption{Lags comparison for sparse arrays.}
\label{fig:Lags_comp}
 \end{figure}

The proposed thinned coprime arrays retain all the properties of conventional coprime arrays, but with $\ceil*{\frac{M}{2}}$ fewer sensors. For comparison, we generate the DOFs including consecutive and unique lags for the sparse arrays under consideration. Unique lags for CADiS with $M' > $ 1 and different cases of $L$ along with thinned coprime array are plotted in Fig. \ref{fig:Lags_comp}(a), while the consecutive lags for nested array, nested CADiS, thinned coprime array and sparsest versions of CADiS are plotted in Fig. \ref{fig:Lags_comp}(b) for an array with fixed number of sensors in the range from 12 to 40 sensors.
\begin{figure*}[htbp]
      \begin{tabular}{>{\centering\arraybackslash} m{0.4 cm} >{\centering\arraybackslash} m{5.35 cm} >{\centering\arraybackslash} m{5.35 cm} >{\centering\arraybackslash} m{5.35 cm}}
     \toprule
    &  (a) Conventional Coprime & (b) Nested  & (c) Thinned Coprime\\
    \hline
      $P$($\theta$)
     &
     \raisebox{-\totalheight}{\includegraphics[width=0.305\textwidth]{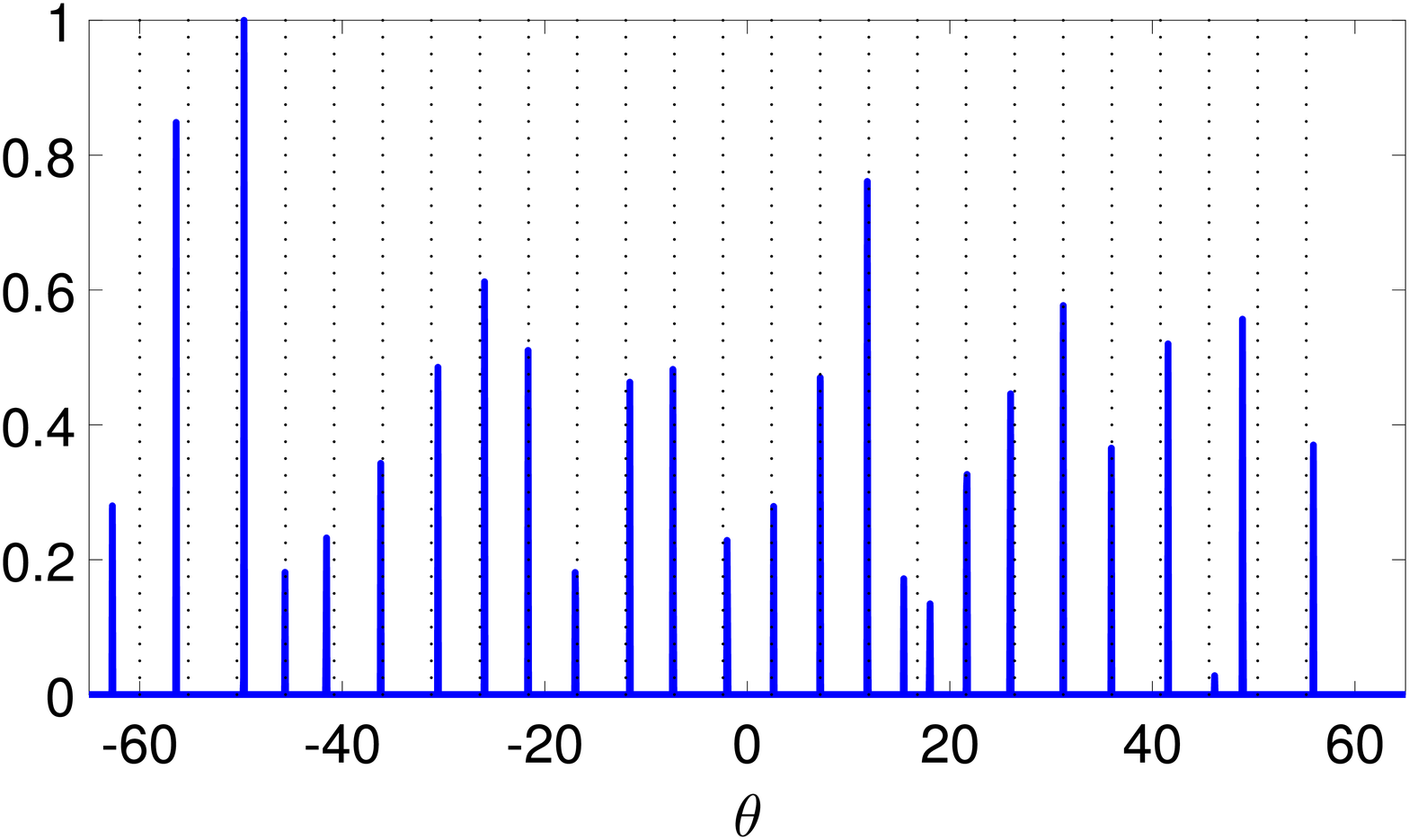}}
      &
     \raisebox{-\totalheight}{\includegraphics[width=0.305\textwidth]{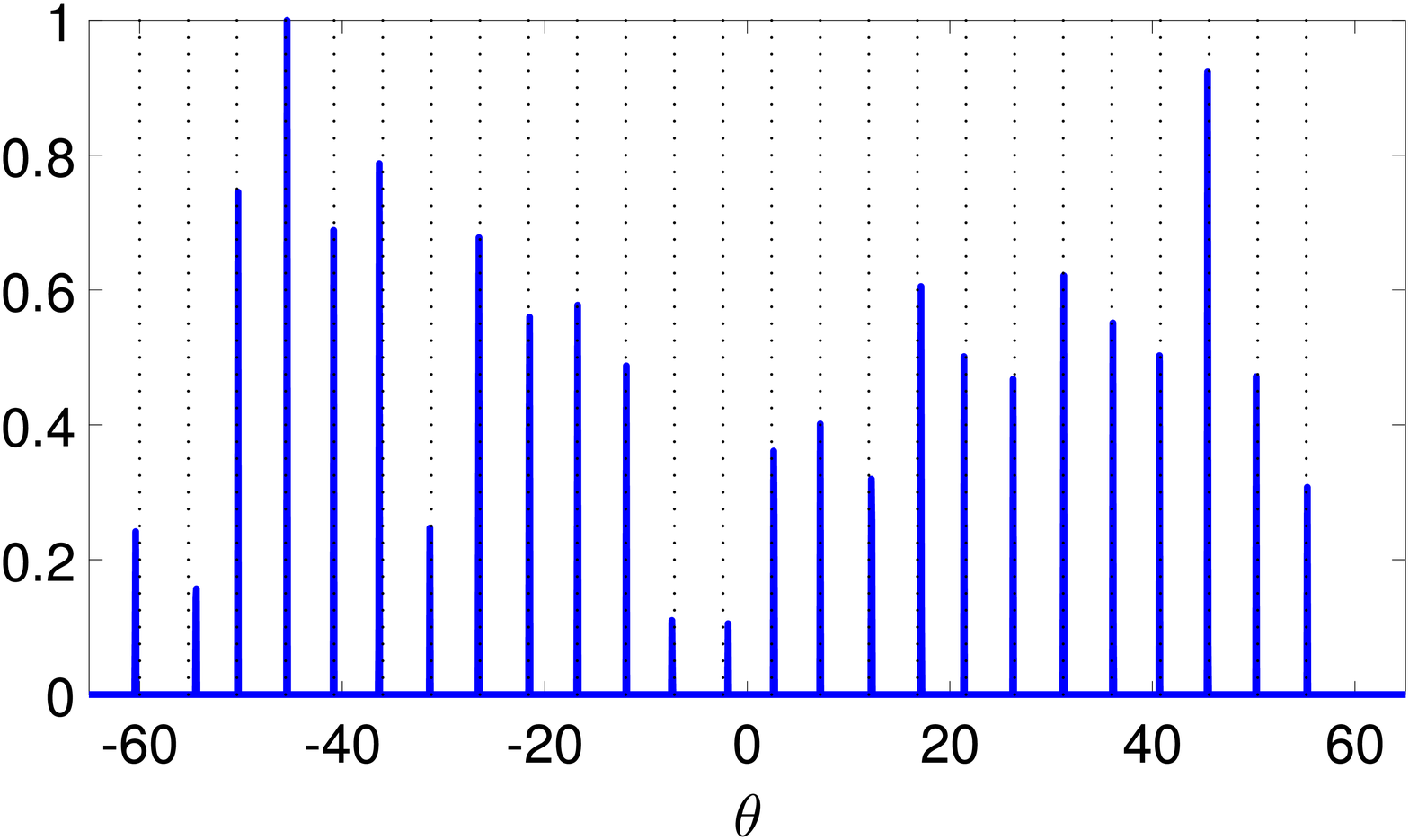}}
      &
     \raisebox{-\totalheight}{\includegraphics[width=0.305\textwidth]{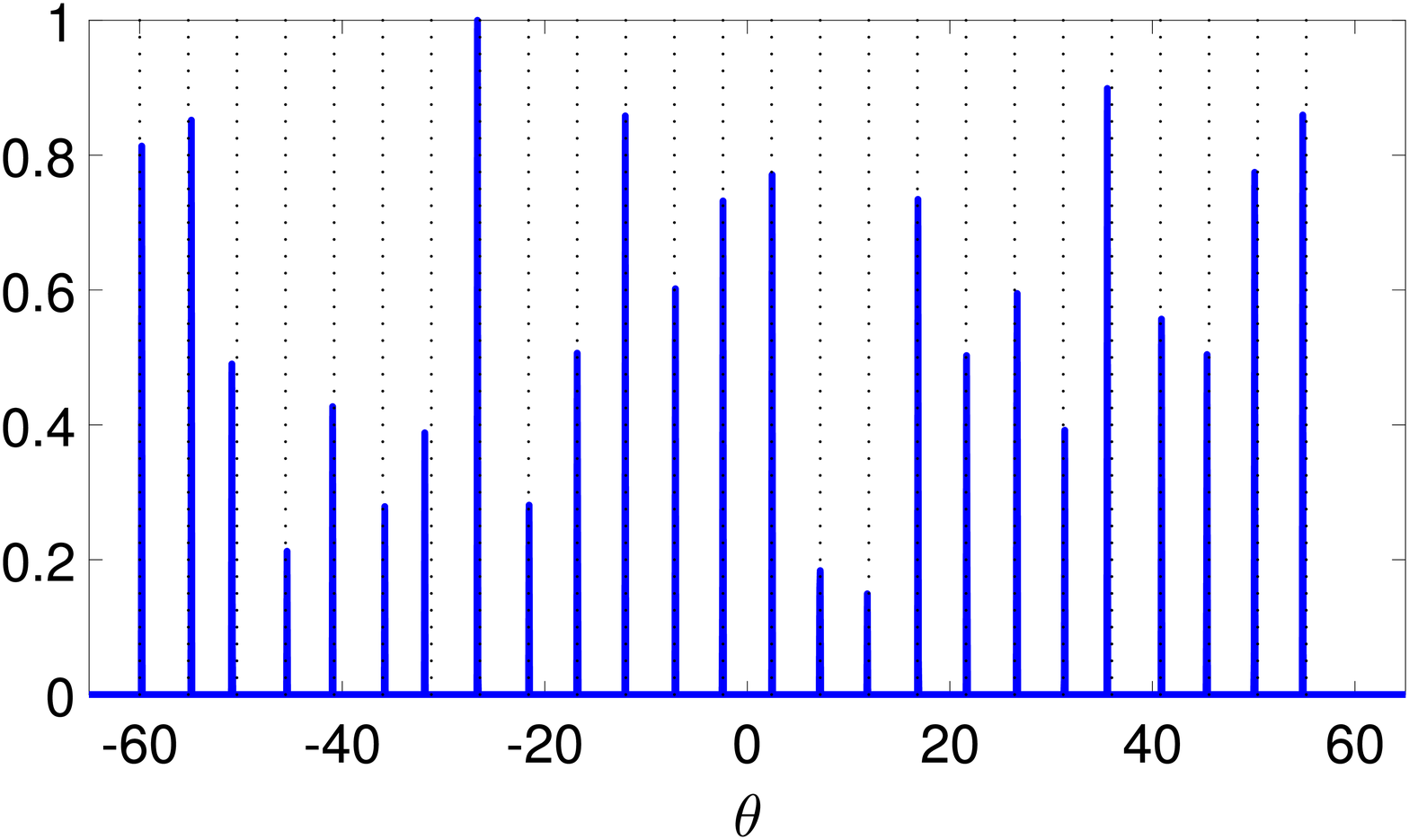}}
     \\ \bottomrule
      \end{tabular}
      \caption{Comparison among (a) Conventional coprime array, (b) nested array, and (c) thinned coprime array for DOA estimation performance. The CS spectrum has been computed with 512 snapshots, 0 dB SNR, 12 sensors and 25 sources marked by dots on the $\theta$ axis.}
      \label{fig:comp}
      \end{figure*}

One potential problem in generating CADiS with $M' > $ 1 for any fixed number of sensors lies in the fact that for most of the cases, $M$ appears to be a prime number thus offering only the possibility of generating nested CADiS with $M' = 1$. For the analysis, all the sparsest versions of CADiS with the maximum possible value of $M'$ less than $M$ have been extracted and their unique and consecutive lags have been calculated. Nested array, nested CADiS and thinned coprime array all can be generated for the considered range of sensors. It can be seen in Fig. \ref{fig:Lags_comp}(a) that the unique lags of thinned coprime array are comparable to the unique lags of the sparsest CADiS with $L = M' + N$, while the sparsest CADiS with $L > N$($M - $2) generates the highest number of unique lags. The unique lags of thinned coprime array in Fig. \ref{fig:Lags_comp}(a) are greater than the hole-free structure of nested array and nested CADiS as depicted in Fig. \ref{fig:Lags_comp}(b). Nested array and nested CADiS produce the highest number of consecutive lags for sparse arrays while the number of consecutive lags for sparsest versions of CADiS in comparison to thinned coprime array, nested array and nested CADiS are very low.

On the whole, sparse versions of CADiS with $M' > $ 1 cannot be generated for an arbitrary number of sensors and possess very low number of consecutive lags to be exploited by MUSIC based DOA estimation methods. Their application lies directly in the CS-based methods, where their unique lags can be utilized. Thinned coprime arrays can be generated for any arbitrary number of sensors. The number of unique lags generated by thinned coprime arrays are much higher than most of the sparse arrays and even the consecutive lags generated by thinned coprime array are on average around 75 percent of the hole-free coarray generated by nested arrays, which proves their application in both MUSIC and CS-based DOA estimation methods. Also the aperture of thinned coprime arrays is found to be on average roughly 1.3 times the aperture of nested arrays.


 \section{Simulation Results with CS-BASED DOA ESTIMATION}
\label{sec:doa}

In this section simulation results for the proposed thinned coprime array are compared with the nested array and the conventional coprime array through the CS-based DOA estimation method.

First we briefly review the CS-based DOA estimation method.  By vectorizing $\textbf{R}_{xx}$ in \eqref{eq:Covar}, we have 

\begin{equation}
\textbf{z}=vec(\textbf{R}_{xx})=\tilde{\textbf{A}}\textbf{b}+\sigma_n^2\tilde{\textbf{I}}=\textbf{B}\textbf{r}
\label{eq:vec}
\end{equation}
where $\tilde{\textbf{A}}=[\tilde{\textbf{a}}(\theta_1),...,\tilde{\textbf{a}}(\theta_Q)]$, $\tilde{\textbf{a}}(\theta_q)=\textbf{a}^{\ast}(\theta_q) \bigotimes \textbf{a}(\theta_q)$, $\textbf{b}=[\sigma_1^2,...,\sigma_Q^2]^T$, $\tilde{\textbf{I}}=vec(\textbf{I}_{S})$. The matrix  $\textbf{I}_{S}$ has a dimension equal to the number of sensors in the sparse array. Additionally, $\textbf{B}=[\tilde{\textbf{A}},\tilde{\textbf{I}}]$ while $\textbf{r}=[\textbf{b}^T, \sigma_n^2]^T=[\sigma_1^2,...,\sigma_Q^2, \sigma_n^2]^T$. Estimating the DOA spectrum of sources $\textbf{r}$  which represents the power of $Q$ sources in addition to the noise power estimate in \eqref{eq:vec} can be achieved by solving the following optimistion problem: 
\begin{equation}
\begin{aligned}
& \text{Min} \quad \|\textbf{r}^{\circ}\|_1
& &\text{s.t.} \quad \|\textbf{z}-\textbf{B}^{\circ}\textbf{r}^{\circ}\|_2 < \epsilon
\end{aligned}
\label{eq:group_sparse}
\end{equation}
where $\textbf{B}^{\circ}$ is a matrix composed of searching steering vectors and $\tilde{\textbf{I}}$, whereas $\textbf{r}^{\circ}$ is a vector of sparse entries to be determined from the search grid. The sensing matrix $\textbf{B}^{\circ}$ and the DOA spectrum estimate vector $\textbf{r}^{\circ}$ are defined over a finite grid $\theta_{1}^{g},...,\theta_{G}^{g}$ where $G \gg Q$. The last entry of $\textbf{r}^{\circ}$ represents the estimate of $\sigma_n^2$, whereas the positions and values of the nonzero entries in other elements of $\textbf{r}^{\circ}$ represent the estimated DOAs and the corresponding signal powers, respectively. 
The value of the threshold $\epsilon$ can be increased to provide more sparsity (less number of nonzero entries) at the cost of increased least square error in the estimates.  The objective function  in \eqref{eq:group_sparse} is convex in $\textbf{r}^{\circ}$ and can be solved using CVX, a software package for specifying and solving convex programs \cite{grant13}.
\begin{figure}[htbp]
\centering
  \includegraphics[width=0.5\textwidth]{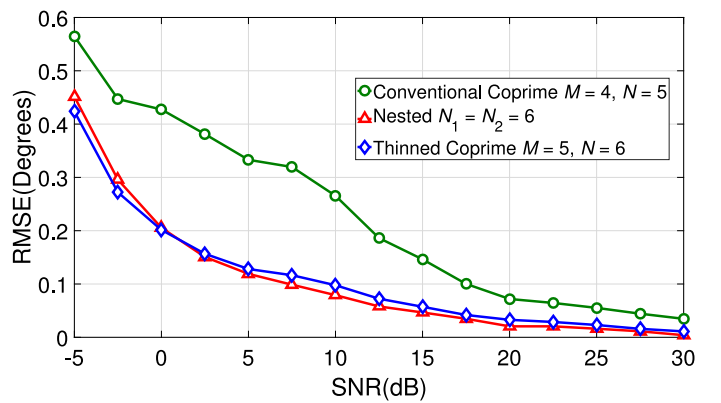}
 \caption{Root mean square error versus input SNR.}
\label{fig:rmse}
\end{figure}

Now we consider a 12-sensor sparse array which are generated with parameters $M=$ 4, $N=$ 5 for conventional coprime array, $N_1=$ 6, $N_2=$ 6 for nested array and $M=$ 5, $N=$ 6 for thinned coprime array. The conventional coprime array has 47 consecutive lags and 59 unique lags, the hole-free nested array has  83 consecutive lags,  and the thinned coprime array has 69 consecutive lags and 89 unique lags. Fig. \ref{fig:comp} represents a normalized CS spectrum $P$($\theta$) for the sparse arrays under consideration. The parameters are 0 dB SNR, 512 snapshots and 25 uncorrelated sources evenly spaced between $-$60$^{\circ}$ and 60$^{\circ}$ with $\epsilon = $ 130 chosen empirically for a clear and fine DOA estimate. A search grid of 3601 angles is formed in the full angle range with a step size of 0.05$^{\circ}$. It can be clearly seen that the conventional coprime array can not detect the 25 sources completely and suffers from false peaks and higher estimation error compared to the  nested array and thinned coprime array.

For a more detailed comparison, we test the three sparse arrays for 20 uncorrelated signals and compute the root mean square error (RMSE) curve against different values of SNR as shown in Fig. \ref{fig:rmse}. Each point on the curve is an average of 500 independent simulation runs and the SNR range is from -5 dB to 30 dB. The value of $\epsilon$ is chosen for the best possible result. It can be seen that the conventional coprime array has a significantly larger estimation error than the nested array and the thinned coprime array, while the latter two perform very much on the same lines especially in low SNR conditions which shows the potential of the proposed thinned coprime array and its improved performance over its parent conventional coprime array structure.

\section{CONCLUSION}
\label{sec:conclusion}
In this paper a so-called thinned coprime array has been proposed, which retains all the properties of the conventional coprime array, but with $\ceil{\frac{M}{2}}$ fewer sensors. For the same number of sensors, they possess greater number of unique lags than the hole-free structure of the nested array and nested CADiS, and comparable number of unique lags to the sparsest CADiS. The consecutive lags of the thinned coprime arrays are around 75 percent to those of nested arrays which showcases their application in both subspace and CS-based DOA estimation methods. Moreover, they can be easily constructed for an arbitrary number of sensors. Simulation results have been provided to show the improved performance by the new structure compared to the conventional coprime array.


\bibliography{mybib}

\end{document}